\documentclass[]{interact}

\usepackage[normalem]{ulem}
\usepackage{multirow}
\usepackage{url}
\usepackage[caption=false]{subfig}
\usepackage{url}
\usepackage{multirow} 
\usepackage[longnamesfirst,sort]{natbib}
\usepackage{quotes} 
\usepackage{bibentry} 
\usepackage{color,soul}
\bibpunct[, ]{(}{)}{;}{a}{,}{,}
\renewcommand\bibfont{\fontsize{10}{12}\selectfont}

\theoremstyle{plain}

\theoremstyle{definition}

\theoremstyle{remark}

\begin{document}

\articletype{ARTICLE TEMPLATE}

\title{Sharing is caring: data sharing in multi-agent supply chains}

\author{
\name{Wan Wang\textsuperscript{a,b}, Haiyan Wang\textsuperscript{a}\thanks{CONTACT Wan Wang.Haiyan Wang.Adam Sobey. Email: ww1a23@soton.ac.uk,
hywang777@whut.edu.cn,
ajs502@soton.ac.uk} and Adam Sobey\textsuperscript{b,c}}
\affil{\textsuperscript{a}School of Transportation and Logistics Engineering, Wuhan University of Technology, No.1178 Heping Avenue Wuchang District, Wuhan, 430063, Hubei, China; \textsuperscript{b}Maritime Engineering, University of Southampton, Burgess Rd,  Southampton, SO17 1BJ, UK;
\textsuperscript{c}Sustainability Mission, The Alan Turing Institute, The British Library, London, NW1 2DB, UK}
}

\maketitle

\begin{abstract}

Modern supply networks are complex interconnected systems. Multi-agent models are increasingly explored to optimise their performance. Most research assumes agents will have full observability of the system by having a single policy represent the agents, which seems unrealistic as this requires companies to share their data. The alternative is to develop a Hidden-Markov Process with separate policies, making the problem challenging to solve. In this paper, we propose a multi-agent system where the factory agent can share information downstream, increasing the observability of the environment. It can choose to share no information, lie, tell the truth or combine these in a mixed strategy. The results show that data sharing can boost the performance, especially when combined with a cooperative reward shaping.  In the high demand scenario there is limited ability to change the strategy and therefore no data sharing approach benefits both agents. However, lying benefits the factory enough for an overall system improvement, although only by a relatively small amount compared to the overall reward. In the low demand scenario, the most successful data sharing is telling the truth which benefits all actors significantly.

\end{abstract}

\begin{keywords}
Multi-Agent Modelling; Supply Chain Management; Hidden-Markov Process; Data Sharing
\end{keywords}

\section{Data sharing in multi-agent supply chain systems}

In many supply chains, the relationship between customer and supplier is close. This creates a symbiotic relationship which requires collaboration. However, each of these companies is still independent and competing for its own profit. These relationships aren't static and their is the potential that these relationships evolve over time, perhaps resulting in a more distant relationship or a former partner moving on to work with a competitor. It creates a challenging dynamic where the companies want the value chain to be successful, especially if no alternative exists, but are seeking to maximise their own returns. 

A key part of building this relationship is data sharing. Data might be shared between supplier and buyer to improve performance across the whole value chain, for example to avoid the bullwhip effect. But it is likely that this is only part of the information as the information holder will only share if they think it benefits them. This results in information asymmetry \citep{chehrazi2025inventory,jannelli2024agentic} which requires complex consensus-seeking \citep{jannelli2024agentic} to reach a near optimal result. This is made even more complex by companies potentially sharing fake information to maximise their profits, which might lead to a loss of the other nodes and where data receivers start to distrust the information received. Trust has been shown to contribute to supply chain cooperation \citep{ozer2011trust, hou2018does} and is therefore important in human relationships, while distrust between corporations can result in an increase in inventory and a decrease in profits \cite{fu2016trust}.

Increasingly, agents are being explored as a method to develop an optimal strategy for supply chain management. Much of the literature focuses on single policy systems where information is already shared across the value chain ~\citep{chen2021data,gijsbrechts2022can,hubbs2020or,keskin2022data,kosasih2022reinforcement,paine2022behaviorally,shi2016nonparametric,stranieri2022deep,stranieri2024performance,vanvuchelen2020use,wang2023agent} but this will be unrealistic for many real-world systems as human developers will not be comfortable with this level of sharing. It seems more likely that each node will develop it's own agent and 
 ~\cite{WANG2025111455}
shows that there can be a large difference between systems where agents are modelled using a single policy, homogeneous, compared to when a separate policy is used for each agent, heterogeneous, It's shown that high demand scenarios exhibit a limited differences in performance, as there are limited strategies available to meet the demand, but in low demand scenarios, the Hidden Decision Markov Process is challenging to solve optimally and there is a $10\times $drop in performance of the agents that each have a separate policy. Those investigating a system of agents with separate policies rarely explore whether the agents can communicate effectively to ensure a stable supply chain ~\citep{ding2022multi,kim2024multi,yu2020multi,sterman2015m,ziegner2025iterative} with only \citep{jannelli2024agentic} investigating LLM agents that can communicate with each other to reach a consensus but that don't share data about their current status.

Therefore, data sharing is explored in this paper through communication of inventory data in a two-echelon supply chain. The factory agent is given the ability to share the truth, a lie \citep{sterman2015m} or no information about their current inventory to the retailer. Each of these scenarios is explored separately and is compared to a scenario where the agent is allowed to select which form of communication that it uses. This is performed across two demands, high and low, as previous research ~\cite{WANG2025111455} shows that there is a change in the power dynamic, with the high demand being factory dominated and the low demand being retailer dominated. Two forms of reward are also explored, a baseline where the agents receive the reward solely for their actions, and a collaborative reward where they are also penalised for stockouts at the other node.

\section{Multi-agent supply chain environment with information sharing 
}
\label{problem}
A two-echelon supply chain is constructed based on \cite{WANG2025111455}. A simple environment allows the effects of communication between agents to be isolated without having to account for emergent behaviours through the supply chain. The environment has two nodes: a factory and a retailer agent; the factory agent buys products from a supplier, attempting to ensure that the inventory contains enough stock to satisfy the retailer, sharing information about it's inventory based on different communication scenarios and setting a price for these products. The retailer attempts to satisfy the customer by ordering enough products from the factory to keep it's inventory above the customer demand. Two customer demand scenarios are tested: one with a high demand $(D \sim Poisson(\mu = 10)$ and one with a low demand $ D \sim Normal(mean=2,std=1))$.


The environment is based on the
Gymnasium APIs framework \citep{brockman2016openai} and Ray's multiagent tools for simulating multi-echelon, multi-agent supply chain environments. Each agent's neural network was built, compiled and trained using Pytorch. All experiments were run on the IRIDIS supercomputer (SLURM, 2023) using CPU Cores Intel(R) Xeon(R) E5-2670, and GPUs (NVIDIA Quadro RTX8000). The multi-agent algorithms are tuned using Ray Tune \citep{moritz2018ray}, a scalable hyperparameter tuning library, which is an open-source library Rllib \citep{liang2018rllib} and Tune \cite{liaw2018tune}. The code is open sourced\footnote{\url{https://github.com/wangwan0910/masc}}.

\begin{figure}[!htbp]
    \centering
\includegraphics[width=2.6in, height=0.7\linewidth]{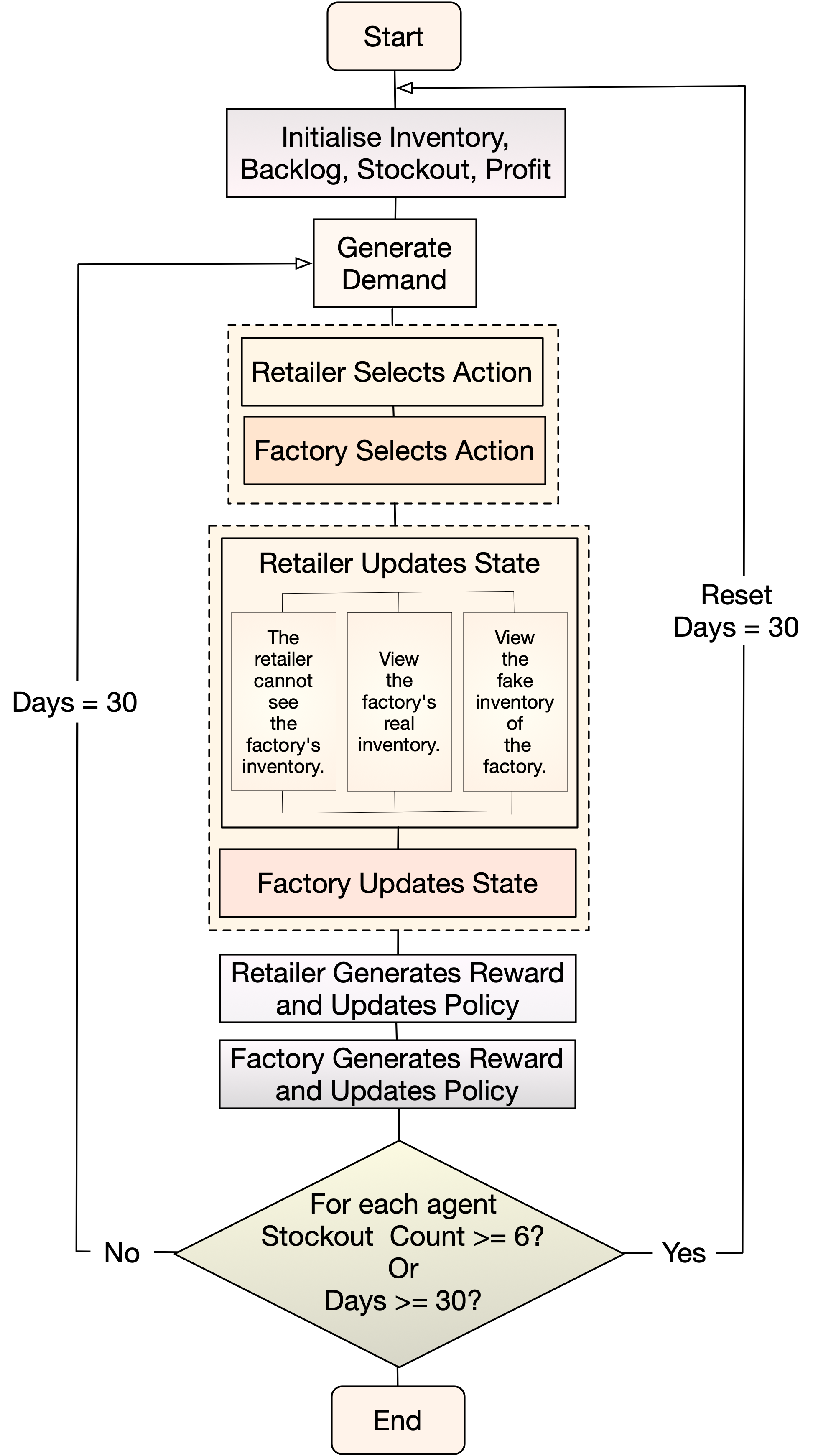}
    \caption{Flow of data sharing through the supply chain environment.}
\label{fig:diagram.}
\end{figure}

Figure. \ref{fig:diagram.} illustrates the flow of goods through the supply chain. Two policies are initialised, the retailer and the factory both have randomised order strategies. The retailer and factory agents then select an action, resulting in inventory being moved between the nodes.
An action is added to the environment in \cite{WANG2025111455} for the factory: it can share its real inventory, fake inventory, or not share any inventory data to the retailer.
Then the agents' states are updated simultaneously, and the retailer's state is updated; the retailer cannot see the factory's inventory, or the retailer can see the factory's inventory, or the retailer sees the factory's fake inventory.
The reward is then calculated and the policy is updated.
For each agent,
it is then determined if the number of stockouts is greater than the maximum allowable value, 6, or whether the number of simulation days exceeds 30 days. If either of these situations occurs, the inventory resets and a new cycle begins.

\subsection{Problem setting}
Both the factory and the retailer agents are represented by a separate SAC agent.  Table \ref{Tab:Hyperpara} documents the hyperparameters used.

\begin{table}[!htb]
\centering					
\caption{Optimal hyperparameter settings for heterogeneous SAC.}				
\label{Tab:Hyperpara}				
\begin{tabular}{ll}
\toprule				
Hyperparameters sets & Values \\ \midrule			
max\_seq\_len&20\\
conv\_activation&relu\\
fcnet\_activation&tanh\\
fcnet\_hiddens&[256,256]  \\
Discount factor (gamma) &0.99\\
Learning rate (lr)&0.001\\
train\_batch\_size&256\\
lstm\_cell\_size&256\\
attention\_dim&64\\
attention\_num\_heads&1\\
attention\_memory\_inference&50\\
attention\_position\_wise\_mlp\_dim &32\\
dim&84\\
initial\_alpha&1\\
tau&0.005\\
prioritized\_replay\_alpha&0.6\\
prioritized\_replay\_beta&0.4\\
prioritized\_replay\_eps&1e-06\\
actor\_learning\_rate&0.0003\\
critic\_learning\_rate&0.0003\\		
preprocessor\_pref      & deepmind                                            \\	
placement\_strategy      & 'PACK'                                      \\	
vf\_share\_layers        & True                                            \\		
attention\_dim          & 64                                               \\	
\bottomrule			
\end{tabular}				
\end{table}

The agents are trained over 60,000 simulated days, at which point the agents have converged to a consistent reward level. This is split into 30 day episodes, $T=30$, after which the environment is reset to the initialisation values and the reward is reset to 0. Each experiment is repeated 10 times. The two agents: the retailer $i=1$, and the factory $i=2$, maximize their rewards by minimising the sum of the inventory and stock-out costs, as shown in Equation (\ref{E2}),  

\begin{equation}
\label{E2}
\begin{aligned}
\centering
\max & \sum_{t=0}^{T}  \begin{bmatrix} 
Sp_{i} \times \sum_{0}^{i}Q_{i,t}-Hc_{i} \times I_{i,t}\\
-Sc_{i} \times  max(Q_{i,t}-I_{i,t},0)-Bc_{i} \times  \\max(I_{i,t} - C_{i},0)\\-c_{i+1} \times Q_{i,t}
\\
\end{bmatrix}\\ \text{subject to:} \\ 
& i \in {1,2}\\
& 0 \le Q_{i,t} \le C_{i}\\
\end{aligned}
\end{equation}

where $Sp_{i}$ is the unit sales price for each agent, $i$, at time $t$; $Q_{i,t}$ is the order quantity; $Hc_{i}$ signifies the inventory holding cost;  $I_{i,t}$ represents the inventory level at each echelon of the supply chain; $Bc_{i}$ is the unit backlog cost; $Sc_{i}$ represents the stock-out cost when the node is out of stock; $C_{i}$ reflects the maximum inventory capacity for each agent and $c_{i}$ reflects the order cost for each agent. Table \ref{Tab:Variables} summarises the environment parameters for the two-echelon supply chain configurations used. 

\begin{table}[!htb]\centering					
	\caption{Environment parameters in the two-echelon supply chain.}				
	\label{Tab:Variables}				
	\begin{tabular}[]{p{1.5cm}p{3.4cm}p{1.2cm}p{1cm}}\toprule				
		\textit{Notation} & \textit{Explanation}& Retailer (i=1)&  Factory (i=2) \\ \midrule			
$Sp_{i}$  & Unit Sales Price& 6& 6 \\	
$Q_{i}$& Order Quantity & [0,20]&[0,20] \\	
$c_{i}$  & Unit Ordering Cost & \textbf{$Sp_{2}$}  &   0.2 \\
$Hc_{i}$ & Unit Inventory Holding Cost &0.2&0.2  \\	
 $I_{i}$ & Initial Inventory Level &10&10 \\	
$C_{i}$ & Inventory Capacity  &19&59  \\
\textbf{$Sc_{i}$} & Unit Stockout Cost &140&70  \\	
$SL_{i}$& Initial Stockout Level & 0&0 \\
 $Bc_{i}$ & Unit Backlog Cost &  1&    1  \\
 $B_{i}$ & Initial Backlog Level  &  0&    0  \\
$D $& Customer Demand &D&$Q_{1}$   \\	
$T$  & Simulation Days &  30 &  30\\
\bottomrule			
\end{tabular}				
\end{table}	



\subsection{State space} 

The state $s_{i,t}$ is defined as a vector defined in Equation (\ref{E:RF}),  

\begin{equation}
\label{E:RF}
s_{i,t}=\left \{ s_{1,t} ,s_{2,t},
\right \}, 
\end{equation}

where the state for the retailer is defined in Equation (\ref{E:R}), where $I_{1,t}$ represents the inventory level at the retailer; $B_{1,t}$ backlog level at the retailer; $S_{1,t}$ represents the stock-out level at the retailer and $D_{1,t}$ is the demand at the customer. 
\begin{equation}
\label{E:R}
s_{1,t}=\left \{I_{1,t},B_{1,t},S_{1,t},D_{1,t} 
\right \},
\end{equation}

The state of the factory is defined by the following Equation (\ref{E:F}), where $I_{2,t}$ represents the inventory level at the factory; $B_{2,t}$ backlog level at the factory; $S_{2,t}$ represents the stock-out level at the factory and $D_{2,t}$ is the demand at the retailer. 

\begin{equation}
\label{E:F}
s_{2,t}=\left \{I_{2,t},B_{2,t} ,S_{2,t},D_{2,t}
\right \}. 
\end{equation}

\subsection{State transition} 

The transition function is implemented according to the material balance constraints in Equation(\ref{E6:transition}) as follows,

\begin{gather}
\label{E6:transition}
\begin{aligned}
I_{i,t+1} &= I_{i,t} + Q_{i,t} - D_{i,t}.\\
\end{aligned}
\end{gather}

The downstream participants' demand is the upstream participants' order. The inventory at the next step is the addition of the current orders to the previous step and the subtraction of the current demand. Backlogs only occur when current inventory is higher than the inventory capacity, and therefore backlogs are always a positive value, which is denoted by Equation (\ref{E6:backlog}), 


\begin{equation}
\label{E6:backlog}
\begin{cases}
I_{i,t} -C_{i} & \text{if } B_{i,t} \text{ is positive} \\
0 & \text{otherwise}
\end{cases}
\end{equation}

Stockouts only occur when current demand/orders are greater than inventory levels and therefore the stockout is always positive, defined as Equation (\ref{E6:stockout}),


\begin{equation}
\label{E6:stockout}
\begin{cases}
Q_{i,t} - I_{i,t} & \text{if } S_{i,t} \text{ is negative} \\
0 & \text{otherwise}
\end{cases}
\end{equation}

\subsection{Action space}
The retailer can make orders from the factory to ensure that the inventory meets the customer demand, with the product price remaining static, the action space for the retailer is expressed in Equation (\ref{ActionR}), 
\begin{equation}
\label{ActionR}
a_{1,t}=\left \{ Q_{1,t}\right \}.
\end{equation}

The factory can make orders from the supplier to ensure its inventory can meet the retailer demand and communicate its stock levels and its communication factor is $\omega_{j,t}$, the action space is expressed in Equation (\ref{E:Base}),

\begin{equation}
\label{E:Base}
a_{2,t}=\left \{ Q_{2,t}, \omega_{j,t} \right \}.
\end{equation}

where $\omega_{j,t}$ represents the different communication scenarios: $j=1$ the No Communication scenario, $j=2$ the Lying scenario and $j=3$ the Trust scenario. In the No Communication scenario, the factory inventory information, $ I_{1,F} = \omega_{1,t} \times I_{2,t}$, is not shared with the retailer and so $\omega_{1,t} = 0 $. In the Lying scenario, the factory provides a fake inventory estimate, $I_{1,F} = \omega_{2,t} \times C_{2}$, where $\omega_{2,t} \in U[0,1)$. In the Trust scenario, the inventory information of the factory, $I_{1,F} = \omega_{3,t} \times I_{2,t}$, is shared completely with the retailer, meaning that $\omega_{3,t} = 1$. In the Mixed scenario, the agent can select from all of the above scenarios: Trust, Lying and No Communication.

\subsection{Observation space}
The retailer has a different observation depending on the scenario. In the No Communication scenario, the retailer has a more limited view of the world with no ability to see the factory's inventory defined in Equation (\ref{RObsNC}), 

\begin{equation}
\label{RObsNC}
obs_{1,t} = \{ \{I_{1,t},B_{1,t},S_{1,t},
D_{1,t}, 0,
p_{t} \}
, t\in \left \{ 1,\cdots T  \right \} \},
\end{equation}

  In this context, $t$ denotes the specific day of $T$, and $p_{t}$ denotes the simulated day. The simulation completes after more than six stockouts or 30 days, so $p_{t}$ is less than 30.  In the other communication scenarios, the inventory communicated to the retailer can be any value from 0 to the maximum inventory level, defined in Equation (\ref{RObs}),

\begin{equation}
\label{RObs}
\begin{aligned}
obs_{1,t} = \{ \{I_{1,t},B_{1,t},S_{1,t},D_{1,t},I_{1,F},p_{t}
\},  I_F \in [0,59),   t\in \left \{ 1,\cdots T  \right \} \}.
\end{aligned}
\end{equation}

where $I_{1,F}$ is the inventory communicated by the factory. In each period $t$, the factory observes the new and previous demand as well as its current state. This observation, $obs_{2,t}$, is defined in Equation (\ref{FactoryObs}), 

\begin{equation}
\label{FactoryObs}
obs_{2,t} = \{ \{I_{2,t},B_{2,t},S_{2,t},
D_{2,t},p_{t} \}, t\in \left \{ 1,\cdots T  \right \} \},
\end{equation}

where $I_{2,t}$ is the inventory , $B_{2,t}$ is the  backlog level, $S_{2,t}$ is the stockout level and $D_{2,t}$ is the demand from the retailer.

\subsection{Reward function}

In the previous literature  ~\cite{WANG2025111455}, reward shaping is introduced where the agents are penalised for the other agent running out of stock. This shows some limited benefits, but it is proposed that part of the lack of success is the lack of observability in the system. This reward is therefore explored in conjunction with the observation space.

\subsubsection{Baseline reward}
In the baseline reward, the agents will learn to choose the optimal action at each state to maximise the agent's profits: 
$r_{i,t}=\left \{ r_{1,t} ,r_{2,t}\right \}$.

 %


   

The reward for the retailer is given in Equation (\ref{BaselineR}),

\begin{equation}
\begin{aligned}
\label{BaselineR}
 \text{$r_{1,t}$} = 
 Sp_{1,t} \times D_{t-1} - 0.2 \times I_{1,t}
   -  Sp_{2} \times Q_{1,t}   \\ - 140 \times \max{(D_{t} - I_{1,t} , 0)  } - 1 \times \max(I_{1,t} - 20, 0),\\
 \end{aligned}
\end{equation}

and the corresponding factory reward is given in Equation (\ref{BaselineF}),
\begin{equation}
\begin{aligned}
\label{BaselineF}
 \text{$r_{2,t}$} =  Sp_{2,t} \times Q_{1,t-1}  -  0.2 \times I_{2,t}- 0.2\times Q_{2,t} \\ - 70 \times \max{(Q_{1,t}- 
 I_{2,t}, 0)} -  1 \times\max(I_{2,t} - 60, 0).
 \end{aligned}
\end{equation}




     


\subsubsection{Collaborative reward}
A reward is taken from \citep{WANG2025111455} to encourage a joint strategy between the agents. Many multi-agent reward schemes use a profit sharing mechanism, which seems unrealsitic in real-world scenarions, and so this sytem provides a penalty for stockouts in the other node as this is a reputational risk for all nodes. This reward is designated Collaboration and defined for the retailer in Equation (\ref{CollaR}),

\begin{equation}
\label{CollaR}
\begin{aligned}
 r_{1,t} -10 \times \max{(Q_{1,t} - 
I_{2,t}, 0) }, 
\\
\end{aligned}
\end{equation}



and for the factory in Equation (\ref{CollaF}),

\begin{equation}
\label{CollaF}
\begin{aligned}
 r_{2,t}  - 20 \times \max{(D_{t} - I_{1,t} , 0)  }
.
\end{aligned}
\end{equation}

\section{Data sharing with 
collaborative reward shaping scenarios
}
\label{shaping}

The supply network is simulated for the three types of communication: Lying, Truth
or Mixed and compared back to the environment with No Communication.
 This
is performed on the high demand and the low demand environments.
As training progresses, there is a notable improvement in the performance of the model, with the strategy reaching a state of equilibrium in the later stages.
The results are
taken from the end of the training period where the agents behaviour is converged to
a single global behaviour.


\subsection{High demand environment with baseline rewards}
In the high demand scenario, the overall reward in the system is highest when the factory always lies about its current inventory, 1335, shown in Table ~\ref{Tab:heter r baseline high}. This is followed by no communication, 1323, with the agents showing that a mixed strategy or a truthful strategy have lower rewards. In the lying strategy the factory is able to make a larger profit than in the other scenarios, 1711, while the retailer makes a similar loss to the no communication and truth scenario. The retailer makes the highest profit in the mixed scenario, -361, although this shows a negative outcome in all scenarios as the price is fixed at a low value. 

\begin{table}[!htb]
    \centering
    \caption{Agent baseline rewards in the high demand environment for the four forms of communication under simulated ten times. Higher scores indicate superior agent performance.}
    \label{Tab:heter r baseline high}
    \begin{tabular}{ccccccc}
        \toprule 
      \textit{Reward} &  \textit{NoComms}  &   \multicolumn{3}{c}{\textit{Communication}}\tabularnewline\cmidrule(l){3-5} &&\textit{Truth}& \textit{Lying}&\textit{Mixed} \\
       \midrule
       Factory   &1695.74   &1,693.40 &1710.62 &1671.95\\
        Retailer    & -373.07 &-376.01&-375.94&-361.12\\
        Global    & 1,322.67 & 1,317.39&1,334.69 &1310.82 \\
        \bottomrule
    \end{tabular}
\end{table}

The mean inventory at the retailer stays at a relatively fixed position with the no communications baseline at the highest level of 14.5 which is similar to the truth, 14.4, and Lying, 14.5, while the mixed scenario shows a lower value of 14.2 which correlates with the improved reward. The agents all receive a similar number of stockouts from 0.059 in the trust scenario, 0.057 in the mixed scenario, 0.046 in the lying scenario and 0.041 in the no communications scenario. This is reversed in the number of backlogs, with the no communications suffering 0.6 and lying 0.59 with truth and mixed at 0.54.

There is more variation in the factory mean inventory, with the higher performing rewards of lying and no communication at a mean value 28, the mixed scenario has a similar mean value of 29 while the trust scenario has a higher a mean of 34. This allows all of the factory agents to avoid stockouts and to generally avoid backlogs except in the lying scenario where 0.001 backlogs were witnessed.

\subsection{High demand environment with collaborative reward shaping}

Rewards are difficult to compare to the baseline agents as in the reward shaping the agents are penalised by the stockouts state of the other agents; the agents will be penalised for its own stockouts and again for the stockouts of the other agent. The reward is therefore adjusted for the following analysis to make it comparable to the baseline results, by adding the reward for the stockouts and backlogs from the opposite node to the final reward.


\begin{table}[!htb]
    \centering
    \caption{The agent receives collaborative rewards for the four forms of communication in the reward shaping high demand  environment under ten  simulations. Higher scores indicate superior agent performance.}
    \label{Tab:heter r high}
    \begin{tabular}{ccccccc}
        \toprule 
      \textit{Reward} &  \textit{NoComms}  &   \multicolumn{3}{c}{\textit{Communication}}\tabularnewline\cmidrule(l){3-5} &&\textit{Truth} &\textit{Lying}& \textit{Mixed}\\
       \midrule
       Factory   & 1653.04 &1635.50&1718.62 &1650.64\\
        Retailer    & -390.74 &-440.39&-406.74&-362.95\\
        Global    &1262.3  &1195.11&1311.88 &1287.69 \\
        \bottomrule
    \end{tabular}
\end{table}

When the agents have a shared reward, the global performance is lower than for the agents with the baseline reward, shown in Table \ref{Tab:heter r high}. The best approach for the global reward is for the factory to lie all the time, 1312. In the collaborative reward scheme, the mixed strategy and no communications are flipped compared to the baseline but the truthful strategy again gives the lowest score, 1195. In the lying strategy, the factory has the highest score, 1719, while the no communications and mixed strategies have a similar reward and telling the truth reduces the factory profit, 1635.50. As with the baseline reward, the retailer performs the best in the mixed scenario, with a reward of -363, although in this case there is a larger difference in performance across the other approaches with the no communications giving a reward of -391 and then a drop to lying of -407 and the truthful scenario giving the lowest score of -440. 

The mean inventory at the retailer again stays at a relatively fixed position which is similar to the baseline reward. The mixed strategy gives a reward of 14.6 followed by lying and truth telling at 14.5 and a drop to no communications at 13.8. This leads to a similar number of stockouts, from 0.059 in the trust scenario, 0.058 in the lying scenario, 0.055 in the mixed scenario and 0.053 in the no communications scenario. This is somewhat reversed in the number of backlogs, with the no communications suffering 0.57, truth 0.54 and lying 0.53, although the mixed scenario gives the lowest number of backlogs 0.46.

There is again more variation in the factory mean inventory, with the higher performing rewards of no communications with 29, lying and mixed with 31 showing lower values than the truth at 35. This allows all of the factory agents to avoid stockouts and to generally avoid backlogs except the truth scenario showing backlogs 1.6\% of the time and the no communications having a backlog 0.001\% of the time.

\subsection{Low demand environment with the baseline reward}

A low demand environment is also considered which in ~\cite{WANG2025111455} is shown to change the power dynamic, with the low demand being retailer dominated. This is also reflected in the current results, with a lower factory reward than the retailer reward which is reversed from the high demand scenario. Although the lower frames leads to a lower reward overall and for each agent, shown in Table ~\ref{Tab:heter r base low}.

\begin{table}[!htb]
    \centering
    \caption{Agent Baseline rewards in the low demand environment for the four forms of communication under 10 times simulation. Higher scores indicate superior agent performance. }
    \label{Tab:heter r base low}
    \begin{tabular}{ccccccc}
        \toprule 
      \textit{Reward} &  \textit{NoComms}  &   \multicolumn{3}{c}{\textit{Communication}}\tabularnewline\cmidrule(l){3-5} &&\textit{Truth}&\textit{Lying}& \textit{Mixed} \\
       \midrule
       Factory &  42.62& 71.82&-5.85&46.27\\
        Retailer    & -94.67 &-190.75& -139.17&-174.39\\
        Global   &  -52.04&-118.92&-145.02&-127.98\\
       \bottomrule
    \end{tabular}
\end{table}

The agents in the low demand environment show a different behaviour to the high demand scenario, with a more significant difference between the rewards when there is and isn't communication, with the best baseline behaviour being no communication with a reward of -52. This is followed by the truth, -119, mixed, -128, and lying, -145. The factory follows a different approach, benefiting most from the truth with a reward of 72, followed by the mixed, 46, and no communications, 43. Finally lying leads to the only negative reward scenario, -6. The retailer in this scenario has a maximum reward with no communication, -95, the same as the global reward,  with the trust scenario giving the lowest reward, -191. This means that despite the high performance of the factory in the trust scenario, the retailers poor performance leads to an overall lower reward. 

The mean inventory at the retailer shows more variation than for the high demand environment with a maximum value of 9 for the no communications environment, and lower values for mixed, 7.5, and lying, 7.3, scenarios, with the lowest values for when the factory is providing a truthful response, 5.5. This leads to a smaller number of stockouts, from 0.008\% in the no communication environment, 0.016\% in the mixed scenario and 0.017\% in the lying scenario with a higher value in the truth telling scenario, 0.024\%, which correlates with the lower inventory value. This is somewhat reversed in the number of backlogs, with the no communication suffering 0.007\%, mixed 0.004\% and truth at 0.001\%. However, the lying scenario has an unusually high level of backlogs, 0.01\% compared to it's mean inventory.

There is less variation in the factory inventory, and a lower mean value, compared to the high demand scenario, with the highest mean value of 23 for the lying scenario. This is followed by similar values of 21 for the truthful communication, 20 for the no communications and 19 for the mixed scenario. Here there are more factory stockouts than for the high demand scenario, the most stockouts occur in the lying scenario, 0.016\%, and then the mixed scenario, 0.008\%, and no communication, 0.007\%, with the lowest number in the truth scenario, 0.001\% There are a similarly low number of backlogs, with the lying scenario having the highest 0.008\%, and the truth scenario, 0.006\% with the no communication, 0.001\%, and mixed scenarios, 0.001\%, at a lower level. 
 
\subsection{Low demand environment with collaborative reward shaping}

The reward shaping is also investigated with the low demand environment, shown in Table \ref{Tab:heter r collab low}. In this scenario the trend is the same as for the baseline low demand, with the most successful strategies being no communications and truth telling but with the benefits of these being amplified. The highest global reward is achieved in the truthful scenario, 22.32. For the global reward the no communications approach gives a reward of 8, with a drop to -128 for the mixed approach and -145 for the lying approach.

\begin{table}[!htb]
    \centering
    \caption{Ten simulated experiments were conducted within a reward-shaping environment to assess the collaborative reward mechanism among agents, specifically addressing low demand across four communication formats. Higher scores indicate superior agent performance.}
    \label{Tab:heter r collab low}
    \begin{tabular}{ccccccc}
        \toprule 
      \textit{Reward} &  \textit{NoComms}  &   \multicolumn{3}{c}{\textit{Communication}}\tabularnewline\cmidrule(l){3-5} &&\textit{Truth} &\textit{Lying}& \textit{mixed} \\
       \midrule
      Factory   & 4.67 & 109.86 &-9.6&33.68\\
        Retailer    & 3.33 & -87.53&-135.17&-88.99\\
        Global   &  8 & 22.32& -144.77&-55.32\\
        \bottomrule
    \end{tabular}
\end{table}

The factory receives similar rewards to the baseline environment with the highest rewards in the truthful scenario, 110, with a drop to the mixed, 34, and no communication, 5, scenario with the lying scenario giving it the worst reward -10. This reflects the same order as for the baseline reward but with a higher reward for the trust scenario and lower values for the other communication approaches. For the retailer, it performs best in the no communication scenario, 3, before a drop to truthful communication, -88, and mixed, -89, with the worst performance also when the information is false, -135. The reward shaping leads to higher rewards for the retailer no matter whether there is communication or not. However, this is large in each case except the lying scenario with only a difference in mean reward of 4. 

In the collaborative scenario the inventory for the retailer shows a reasonable drop in mean value, with a maximum value again in the no communications environment of 6, while the truthful value has a value of 5 and the mixed and lying scenarios having a lower value at 4.  This leads to a similar number of stockouts, from 0.017\% for the lying and the truthful communication, 0.013 \% for the mixed communication and 0.009 \% for the no communications but very few backlogs with a small number in the no communications scenario 0.002\% and none in the other scenarios. 

The factory inventory is similar to the baseline reward that there is not much variation. The highest mean value of 18 is seen in the no communication scenario, 17 for the lying scenario, 15 for the truthful communication and 14 for the mixed scenario. This results in fewer factory stockouts, with the lying scenario having the most at 0.004\%, 0.003\% for the mixed scenario, 0.001\% in the no communication scenario and 0 for the truthful scenario. For the backlogs, there are again relatively few backlogs with the lying scenario showing 0.048\% and truthful communication resulting in 0.009\% while the mixed scenario shows no backlog incidents. However, the no communication scenario shows a high level of these incidents, 0.32\%.

\section{Comparison of data sharing strategies}

The data sharing between the agents makes a relatively small difference in the high demand scenario, between -9.5\% to 0.9\%. This is similar to the difference in performance between homogeneous and heterogeneous systems in \cite{WANG2025111455}, where the homogeneous systems will automatically share knowledge between systems. This is likely due to the high demand which requires  a large inventory to ensure that the demand can be met with little room for a change in strategy. However, there is a bigger difference in the low demand scenario, where there is more choice around the strategies to meet the demand, 142\% increase, with both factory and retailer benefitting. Again, showing similarities to \cite{WANG2025111455}, where there is a large drop in performance for the heterogeneous agents compared to the homogeneous ones and where there seems to be more importance to data sharing. 

\begin{figure}[!htbp]
    \centering
\includegraphics[width=4.6in, height=0.49\linewidth]{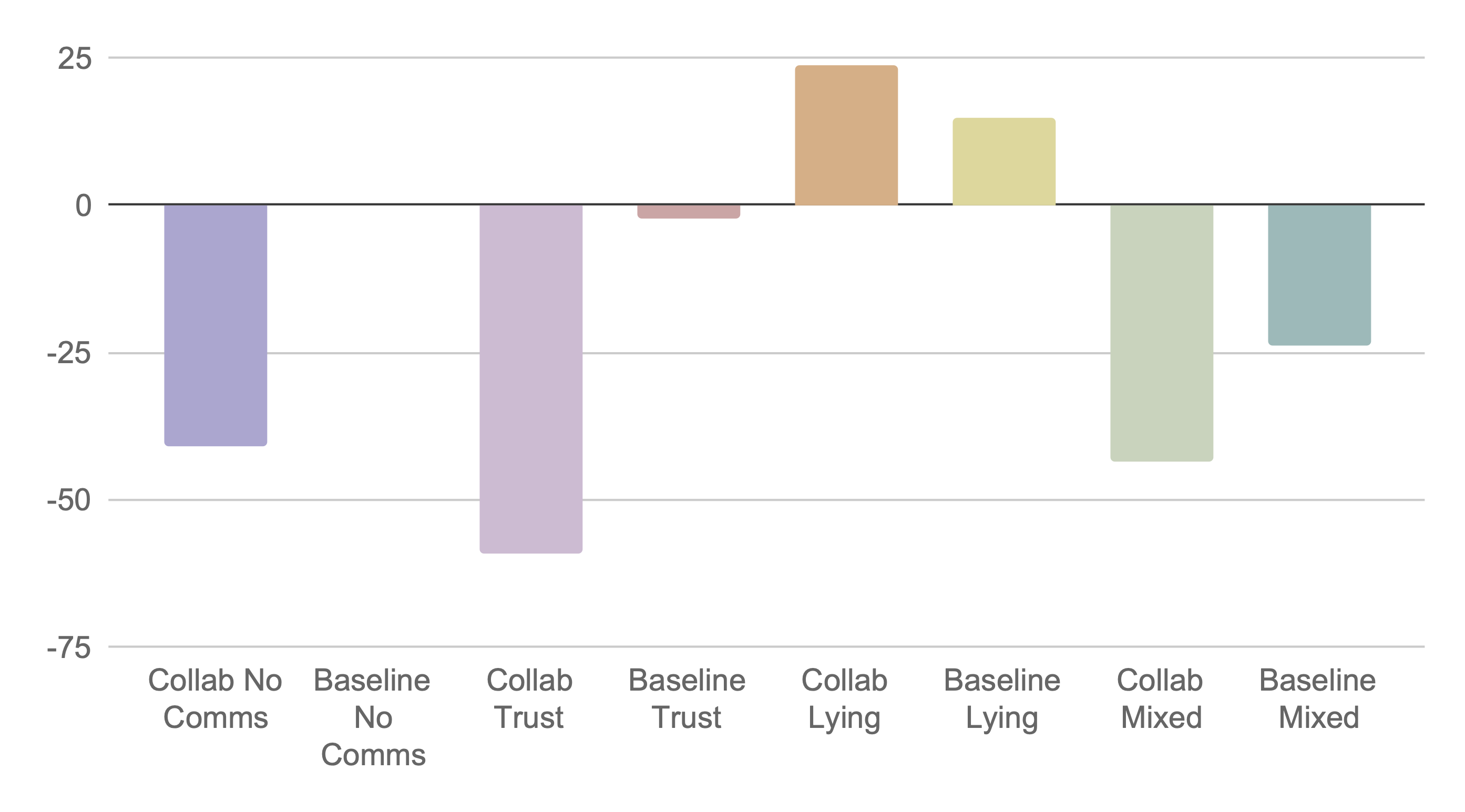}
    \caption{Comparison of factory rewards for the various communication approaches to the baseline environment with No Communication in the high demand environment.}
\label{fig:Comparisonhigh.}
\end{figure}

A comparison of the reward for the factory in the high demand scenario is shown in Figure \ref{fig:Comparisonhigh.} where the only data sharing that outperforms the baseline is for the factory to lie about it's inventory, this is higher in the collaborative reward. It suffers negative rewards in both mixed scenarios and  trust scenarios, although this is small for the trust baseline reward. 

\begin{figure}[!htbp]
    \centering
\includegraphics[width=4.9in, height=0.46\linewidth]{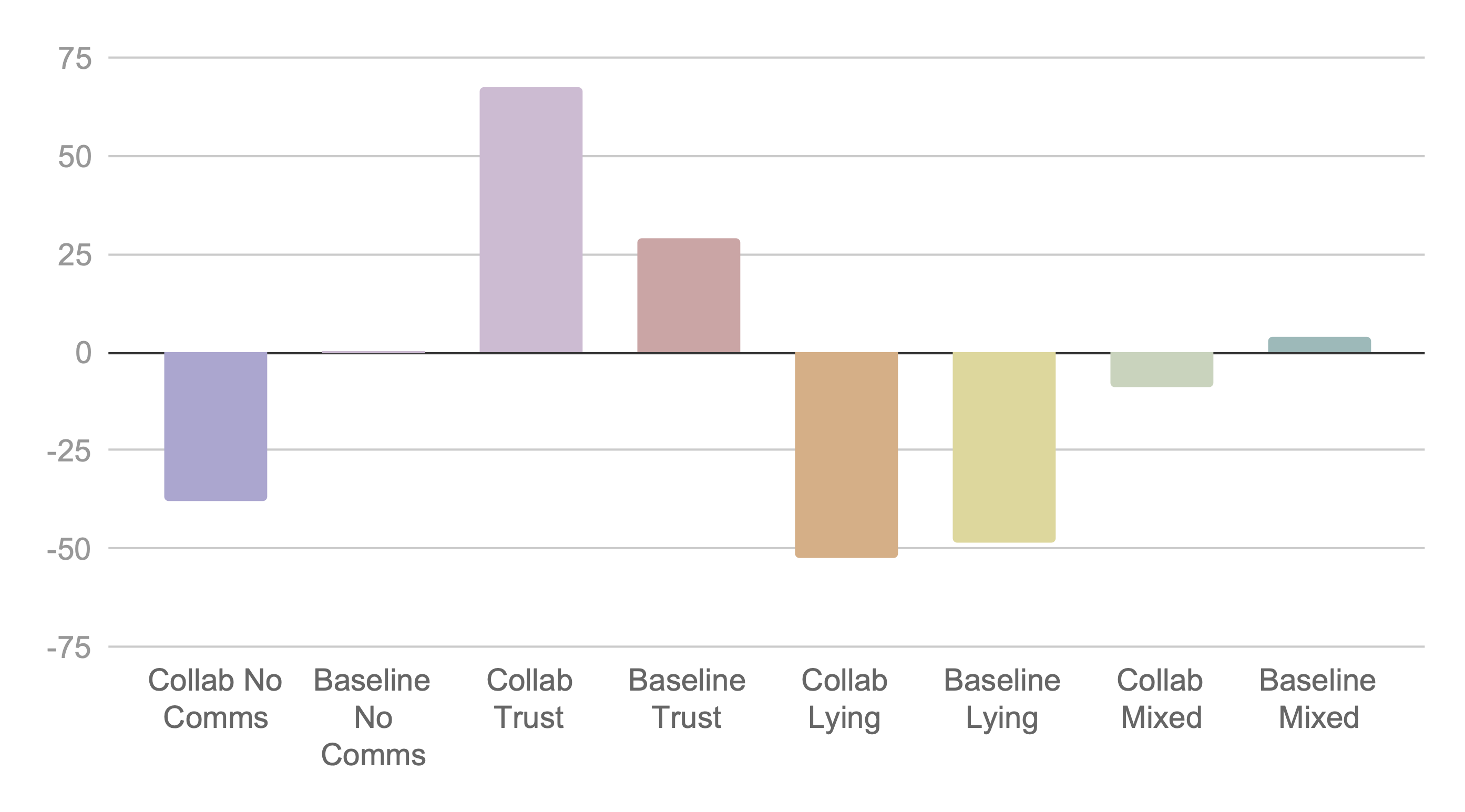}
    \caption{Comparison of factory rewards for the various communication approaches to the baseline environment with No Communication in the low demand environment.}
\label{fig: Comparisonlow.}
\end{figure}

This is reversed in the low demand scenario, shown in Figure \ref{fig: Comparisonlow.}, where the top performance comes when the factory tells the truth about it's inventory, this is higher with the collaborative reward. There is also a small benefit in the mixed communication strategy with the baseline reward, although this is a small detriment with the collaborative reward. Lying always penalises the factory when there is a low demand. 

\begin{figure}[!htbp]
    \centering
\includegraphics[width=4.9in, height=0.46\linewidth]{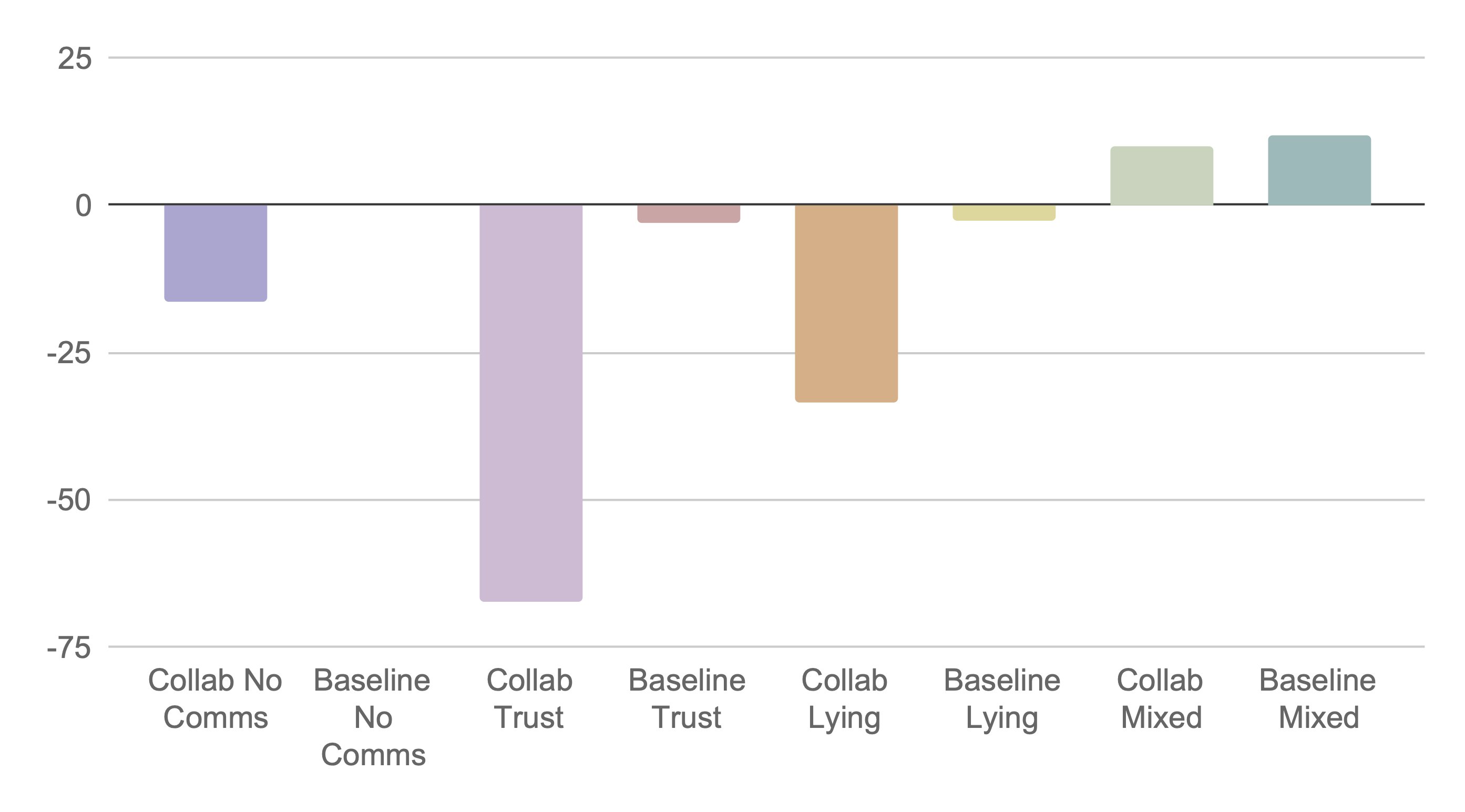}
    \caption{Comparison of retailer rewards for the various communication approaches to the baseline environment with no communication in the high demand environment.}
\label{fig: Comparison high retailer.}
\end{figure}

The retailer only improves it's reward in the mixed data sharing scenario, as depicted in Figure \ref{fig: Comparison high retailer.}. In contrast, the other forms of data sharing show a negative reward with small loses for the baseline lying and trust. As with the factory reward, there are larger negative rewards in the collaborative scenarios.

\begin{figure}[!htbp]
    \centering
\includegraphics[width=4.9in, height=0.46\linewidth]{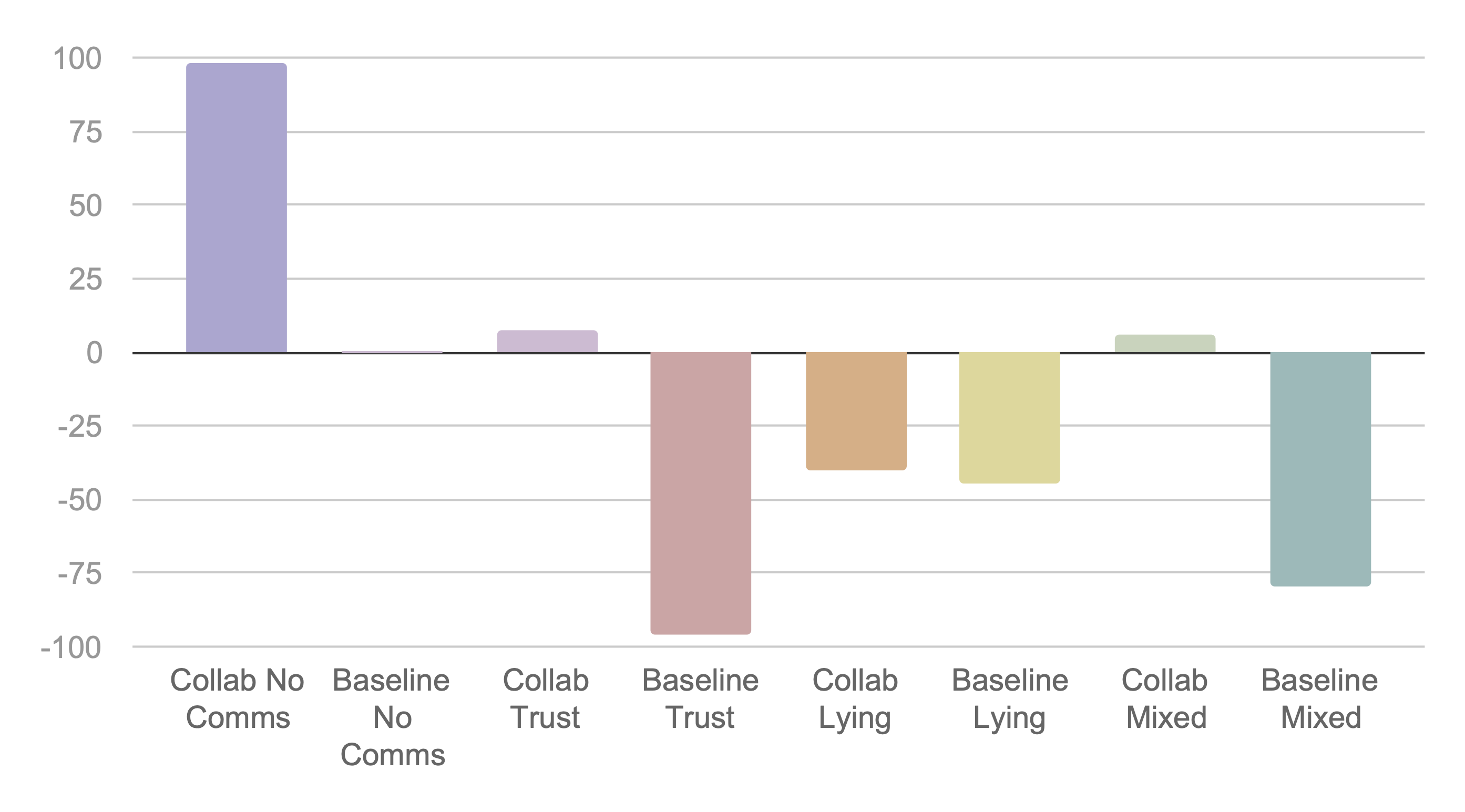}
    \caption{Comparison of retailer rewards to the baseline environment with no communication in the low demand environment.}
\label{fig: Comparison low retailer.}
\end{figure}

In the low demand, the retailer is most successful in the no communication collaboration scenario. Mixed and trust collaboration scenarios earn slightly more rewards than the baseline with no communication, but have negative outcomes in other environments. This is shown in Figure \ref{fig: Comparison low retailer.}.

\subsection{Collaborative reward shaping}

Collaborative reward shaping is shown to be beneficial to the agents and the system when in combination with the correct type of communication but also leads to the largest losses when it is used in connection with the worst communication approaches. 

\begin{figure}[!htbp]
    \centering
\includegraphics[width=2.3in, height=0.3\linewidth]{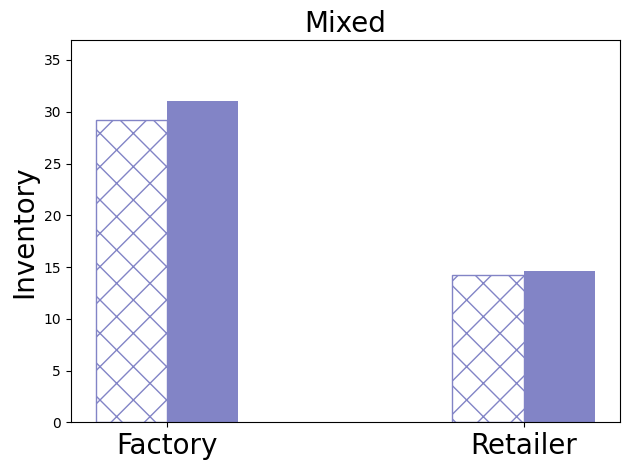}
\includegraphics[width=2.3in, height=0.3\linewidth]{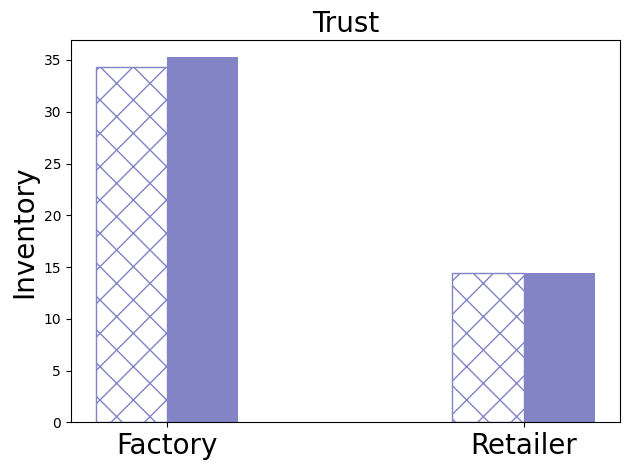}
\includegraphics[width=2.3in, height=0.3\linewidth]{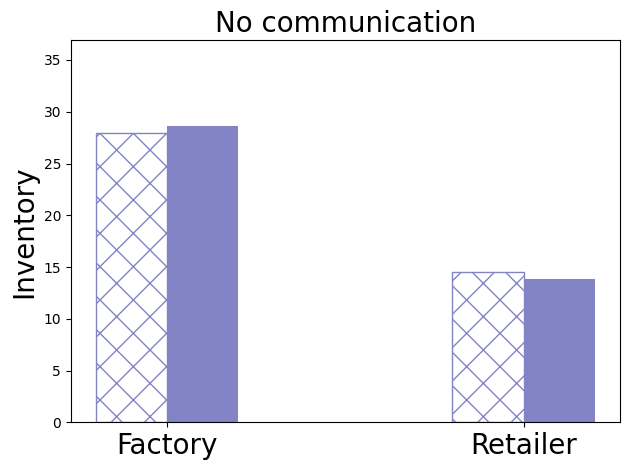}
\includegraphics[width=2.3in, height=0.3\linewidth]{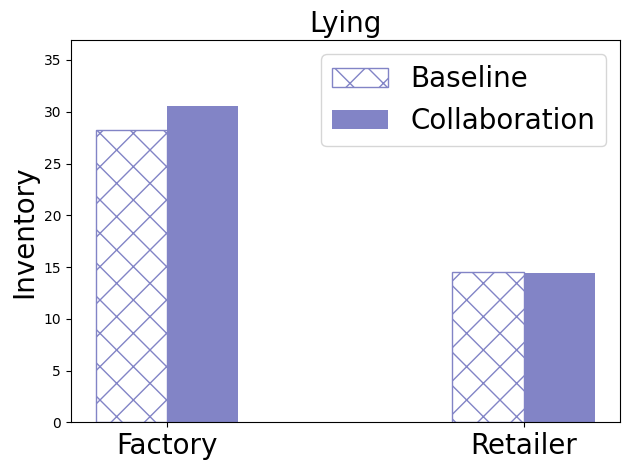}
    \caption{A comparison of inventory levels under the baseline and collaborative scenarios for the high demand. }
\label{fig: Comparison high full.}
\end{figure}

In the high demand scenarios, the reward shaping negatively effects the total reward in the system. In the trust and mixed scenarios the collaborative reward increases the losses for the factory. While for the retailer, it is the trust and lying scenarios. The increase reward for the mixed scenario is less than the baseline, and not as much as the reduction in performance for the factory. The collaboration reward shows a slight increase in the average inventory but this is small, 31.4 compared to a baseline average of 29.9 for the factory, shown in Figure \ref{fig: Comparison high full.}. There are no stockouts in this scenario and so there are no benefits to the collaborative reward but the trust and no communications scenarios start to receive backlogs, as does the baseline lying. For the retailer the inventory is also higher, 10.41 compared to a baseline average of 10.38 for the retailer. This results in a higher number of stockouts, 0.051 for the retailer compared to a baseline of 0.056 for the collaborative reward. However, the number of backlogs is lower, 0.52 compared to 0.56 for the baseline. 

\begin{figure}[!htbp]
    \centering
\includegraphics[width=2.3in, height=0.3\linewidth]{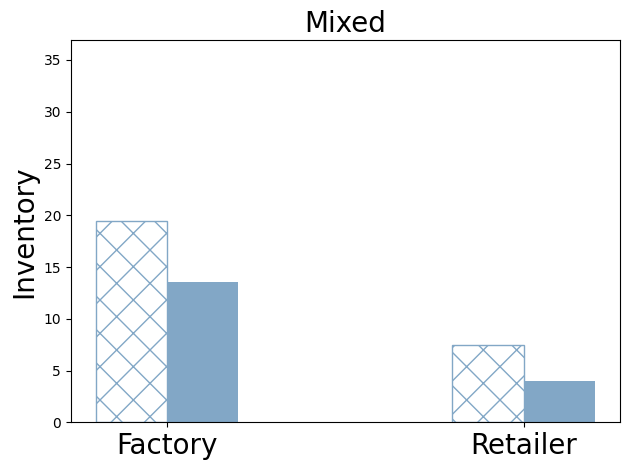}
\includegraphics[width=2.3in, height=0.3\linewidth]{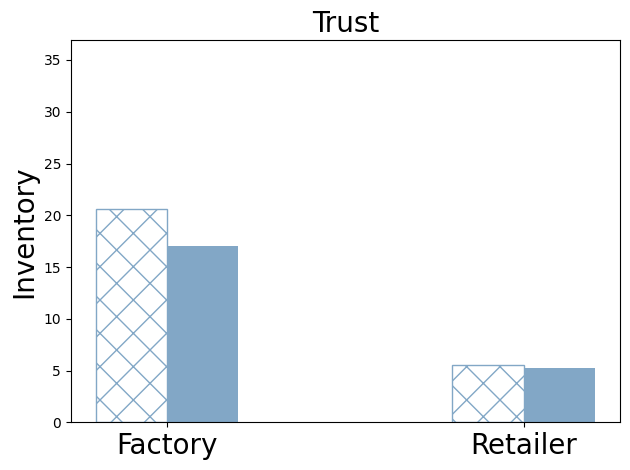}
\includegraphics[width=2.3in, height=0.3\linewidth]{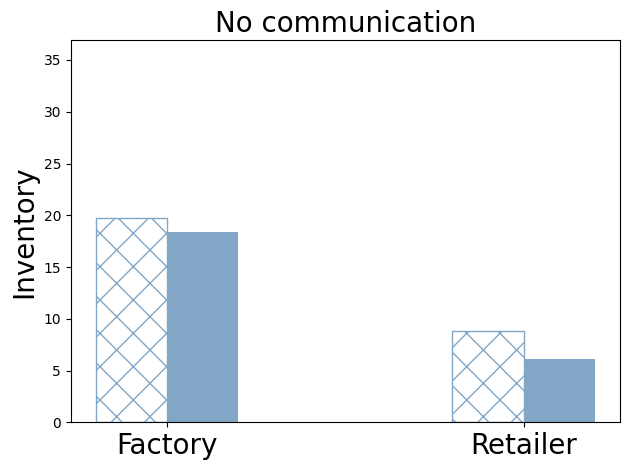}
\includegraphics[width=2.3in, height=0.3\linewidth]{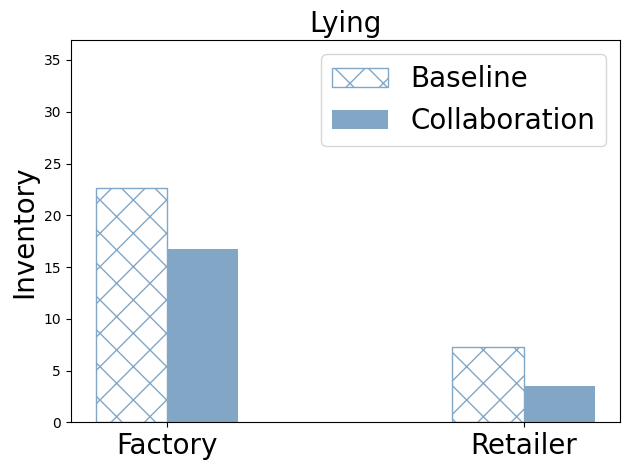}
    \caption{A comparison of inventory levels under the baseline and collaborative scenarios for the low demand.}
\label{fig: Comparison low full.}
\end{figure}

 In the low demand scenario, the inventory is much lower for the collaborative environment with an average of 16 for the factory compared to a baseline of 20.6 for the baseline and 4.66 for the retailer compared to a baseline of 7.3 for the baseline which is shown in Figure \ref{fig: Comparison low full.}. Despite this the number of retailer stockouts and backlogs stay relatively similar, with a stockouts of 0.014 compared to 0.016 for the baseline and 0.0005 backlogs compared to 0.006 for the baseline. This varies more for the factory, with the stockouts at 0.002 compared to 0.007 for the baseline and 0.09 for the backlog compared to 0.004 for the baseline. The trust inventory is an outlier for the baseline environment, 5.5 for the retailer which is the lowest of the baseline inventories and 20.6, which is the average for the factory. While the retailer keeps a higher inventory, 5, in the collaborative trust scenario and this enables a lower value at the factory, 15.4. 

\subsection{
The madman strategy
}

In the mixed scenario, agents often find it difficult to identify a strategy that offers a reward as high as the other communication methods. The breakdown of the communication is given in Table \ref{Tab:percent BC high}. Even though the lying scenario is best for the high demand scenario for the factory and trust is best for the low demand, the agents are not able to find this scenario. The agents use a strategy that replicates the madman strategy \citep{Madman}, where irrational behaviour is used to confuse the opposition in coercive bargaining and has seen benefits under certain conditions \citep{McManus:2019}. The factory shifts away from using trust in the baseline high demand scenario and towards the no communication and lying strategies. This is supported by the fact that lying is the most effective strategy for the factory. However, it is not able to reach a consistent strategy. In the collaborative environment, the agents have a small shift away from trust but the strategy is fairly spread across the different communication approaches. In the low demand scenario, the baseline strategy again has a small shift towards lying which is not an effective strategy for the agent. However, for the collaborative reward the agent is able to reduce the percentage of lying performed. The average factory inventory for the mixed strategy is 16.5 compared to 19 for the other strategies, which is replicated in the retailer with a mixed inventory of 5.7 compared to 6 for the other strategies. This mildly reduces the stockouts and the backlogs for the retailer, which is where the reward is lowest. However, for the factory this increased the number of stockouts, where is performs worse than the lying strategy.

\begin{table}[!htb]
    \centering
        \caption{Mixture of communication styles for the mixed environments.}
    \label{Tab:percent BC high}
    \begin{tabular}{ccccccc}
        \toprule 
        & \textit{Trust} & \textit{NoComms}  & \textit{Lying}& Total Reward  \\
        \midrule
        High Demand &  &   & &   \\
        \midrule
        Baseline  &19\% &39.7\% & 41.3\% & 1310.82  \\
        Collab &  25.5\% & 38.62\% &36\%  & 1287.69\\
             \midrule
        Low Demand &  &  & &   \\
        \midrule
        Baseline  & 29.2\%  &31\%  &40.16\%  &-127.98  \\
        Collab &30.8\%  &44.04\%  &25.16\%&-55.32 \\
        \bottomrule
    \end{tabular}
\end{table}

\section{Discussion and limitations}
\label{Discussion}

The results show that there is no consistent form of data sharing that works best in all scenarios. The retailer might be expected to benefit from the truth most of all. In the collaborative scenario, where it is penalised for factory stockouts, it should reduce it's orders and this is what is seen. However, even in the baseline it needs to ensure there is enough stock when it orders and so there is a driver towards a more collaborative strategy. This isn't reflected in the high demand scenario. However, the difference in reward is lower due to the need to keep inventories so high and it seems there is less room to adapt strategies. The madman strategy strangely seems to benefit the retailer but not the factory. It seems that finding the best form of data sharing is a challenging task for the agents and needs to have additional reward shaping or architectural changes to support it. The retailer demonstrates a robustness to the tactic of miscommunicating the stock, and it seems that while the factory can at times use deception to improve its reward, this doesn't automatically have a large detrimental effect on the retailer. In these scenarios, the reward shaping is more effective than previously proposed by  ~\cite{WANG2025111455}. When the agents are in an environment where it is easier to meet the demand, they can collaborate more effectively when they have access to more information about their environments and we demonstrate that sharing is caring.

In all experiments, including different levels of communication and different reward shaping, the factory agent incurs much fewer stockouts and backlogs than the retailer agent, which may be related to the environmental setup, where the two echelon supply chain turns over quickly in the absence of lead times for delivery. If the upstream supplier consistently meets a factory's demand, the factory agent are better able to avoid stockouts and backlog but they have a larger potential inventory and use this to avoid stockouts. The retailer with a small storage is at a greater risk of a stockouts. Further experiments should explore a longer chain of agents, but also looking to see if having multiple upstream competitors changes the type and the effectiveness of communication. Sharing information and coordinating actions among actors in model training is essential for improving the overall performance of a supply chain when applying heterogeneous deep reinforcement learning. In addition, the lying performed in these experiments is applied in a simple manner, in that it is a random value. A more strategic form of lying could be devised that would allow the factory a greater chance of gaining a reward.

\section{Conclusion}\label{Conclusion}

Agents are increasingly explored to help to optimise inventory management in supply networks. Data sharing is an important topic within the inventory management literature, indicating that sharing more data leads to more efficient supply chains. However, in multi-agent modelling of supply chains, the effects of data sharing have not been explored as agents can only observe their own inventory. There is the potential that this communication can be used to benefit the system but also the potential that the designers/agents might lie
to gain an advantage. In this paper, a high demand and a low demand environment are used, as previous research shows a change in the power dynamic within these supply chains, and with a baseline reward where the agent's reward is solely based on their performance and a collaborative reward, where the agents are penalised for stockouts at the other node. The results show that the optimal style of data sharing is different between the high and low demand scenarios. Data sharing is shown to boost the performance of the system when aligned with the correct communication strategy, but can be detrimental to the system when the incorrect strategy is used. In particular, the data sharing is shown to boost the performance of the cooperation reward shaping, making it more effective with the correct data sharing but also suffering larger losses when the wrong type of data sharing is used. In the high demand scenario, where there is a limited ability to change strategies, the most successful strategy is lying, which gives a small benefit to the factory, 0.9\%, which is more than the detriment to the retailer, -0.7\%; this leads to a small benefit to the system overall. There is no strategy that benefits both agents in this demand. In the low demand scenarios, where the capacity of storage is substantially higher than the demand, the truth is the best policy, increasing the performance of the factory, 158\%, and the retailer, 7.5\%. For multi-agent supply chain agents, sharing is caring.


\section*{CRediT}
Wan Wang Conceptualization; Funding acquisition; Investigation; Methodology; Project administration; Validation; Visualization; Roles/Writing - original draft; and Writing - review \& editing
Haiyan  Funding acquisition; Project administration; Supervision; Writing - review \& editing
Adam Conceptualization; Methodology and Writing - review \& editing

\section*{Declaration of competing interest}
The authors declare that they have no known competing financial interests or personal relationships that could have appeared to influence the work reported in this paper.

\section*{Acknowledgments}
The authors acknowledge the use of the IRIDIS High-Performance Computing Facility, and associated support services at the University of Southampton, in the completion of this work. Wan Wang is funded by the China Scholarship Council. We would like to thank Lloyd’s Register Foundation for their support during this research. Any errors or discrepancies are our own responsibility.

\bibliographystyle{plainnat}
\bibliography{interactapasample}








\appendix
\newpage

\end{document}